# Directional perfect absorption using deep subwavelength low permittivity films

**Ting S. Luk,[1,2,#,*] Salvatore Campione,[1,2,#,$] Iltai Kim,[1,2] Simin Feng,[3] Young Chul Jun[4], Sheng Liu,[1,2] Jeremy B. Wright[1,2], Igal Brener[1,2], Peter B. Catrysse[5], Shanhui Fan[5], Michael B. Sinclair[1]**

[1]*Sandia National Laboratories, P.O. Box 5800, Albuquerque, NM 87185, USA*
[2]*Center for Integrated Nanotechnologies, Sandia National Laboratories, P.O. Box 5800, Albuquerque, NM 87185, USA*
[3]*Naval Surface Warfare Center Dahlgren Division, 18444 Frontage Road, Dahlgren, VA 22448*
[4]*Department of Physics, Inha University, Incheon 402-751, Republic of Korea*
[5]*Edward L. Ginzton Laboratory, Stanford University, Stanford, California 94305-4088, USA*
[#]*These authors contributed equally to this work and are joint first authors.*
[*] *tsluk@sandia.gov ;* [$]*sncampi@sandia.gov*

## Abstract
We experimentally demonstrate single beam directional perfect absorption (to within experimental accuracy) of *p*-polarized light in the near-infrared using unpatterned, deep subwavelength films of indium tin oxide (ITO) on Ag. The experimental perfect absorption occurs slightly above the epsilon-near-zero (ENZ) frequency of ITO where the permittivity is less than one. Remarkably, we obtain perfect absorption for films whose thickness is as low as ~1/50$^{th}$ of the operating free-space wavelength and whose single pass attenuation is only ~ 5%. We further derive simple analytical conditions for perfect absorption in the subwavelength-film regime that reveal the constraints that the ITO permittivity must satisfy if perfect absorption is to be achieved. Then, to get a physical insight on the perfect absorption properties, we analyze the eigenmodes of the layered structure by computing both the real-frequency/complex-wavenumber and the complex-frequency/real-wavenumber modal dispersion diagrams. These analyses allow us to attribute the experimental perfect absorption behavior to the crossover between bound and leaky behavior of one eigenmode of the layered structure. Both modal methods show that perfect absorption occurs at a frequency slightly larger than the ENZ frequency, in agreement with experimental results, and both methods predict a second perfect absorption condition at higher frequencies attributed to another crossover between bound and leaky behavior of the same eigenmode. Our results greatly expand the list of materials that can be considered for use as ultrathin perfect absorbers and also provide a methodology for the design of absorber systems at any desired frequency.

## 1. Introduction

For many applications, such as energy conversion, optical modulation, spectral filtering, and sensing, it is desirable to maximize the efficiency of photon absorption for specific conditions of wavelength and incidence angle. Furthermore, the ability to absorb a large fraction of an incoming optical field in an extremely thin film is beneficial for processes such as photovoltaic energy conversion where carrier extraction efficiency is better for thinner films, or for devices such as bolometers that require large absorption with small thermal mass. One approach that has been successful in this regard is the use of structured materials such as diffraction gratings and metamaterials [1-10]. In these systems, absorption enhancements arise from the enhanced electromagnetic near-fields that occur at the resonant frequencies of the structures. Another approach that has been investigated is based on critical coupling to the surface plasmon modes of planar metal-dielectric-metal structures [11, 12]. A much less explored strategy relies on tailoring stratified dielectric media, in which at least one layer exhibits some loss, to exhibit *perfect absorption* (PA), a process in which (analytically) 100% of light is absorbed in a medium. Such a possibility for wavelength-scale-thick films was recognized quite some time ago, with a number of theoretical papers [13-15] and one experimental work [16] devoted to the elucidation of this process (though not termed perfect absorption in these works). Recently, the perfect absorption concept was explored in the context of laser cavities leading to the so-called "coherent perfect absorption" [17, 18]. Still more recently, the intriguing possibility that perfect absorption can be achieved using deep subwavelength films of highly lossy, high permittivity materials on a variety of substrates has been proposed and experimentally verified [19-21]. It is thus of interest to inquire under what conditions and for what materials similar results can be achieved with deep subwavelength films of low permittivity materials. Although not discussing perfect absorption, the theoretical results of Godwin and Mueller [22] pointed to the possibility that perfect absorption could be achieved near the plasma frequency of a layer of diluted plasma backed by a metal. Similarly, Harbecke et al. [23] theoretically showed strong absorption above the $SiO_2$ longitudinal optical phonon frequency in metal backed films. In addition, a theoretical analysis of perfect absorption in the context of an anisotropic epsilon-near-zero (ENZ) metamaterial has recently appeared [24].

In this paper, we present in Sec. 2 an experimental demonstration of perfect absorption for deep subwavelength films of indium tin oxide (ITO) on metal substrates (Ag) at frequencies just above the ENZ frequency where both the real and imaginary parts of the ITO permittivity are less than one. We find that the incidence angle for perfect absorption increases as the ITO thickness decreases, and that perfect absorption can be obtained even when the ITO thickness is decreased to 24 nm (~ 1/50$^{th}$ of the free space wavelength of 1.1 µm). We present a simple theoretical analysis of the PA process in Sec. 3, obtained within the perfect electrical conductor approximation for the metal substrate, which leads to a set of design equations highlighting the constraints that the ITO permittivity must satisfy if perfect absorption is to be obtained in the subwavelength limit. These equations show that as the real part of the permittivity of the thin layer approaches zero, the film thickness required for PA *decreases* with decreasing loss in the film, in stark contrast with conventional absorbers which require larger thicknesses as the material loss is decreased, but in agreement with previous investigations dealing with the well-known ENZ perfect absorption [23]. We then analyze in Secs. 4 and 5 the behavior of the electromagnetic eigenmodes of the layered structure and compute the modal dispersion diagrams for both the real-frequency/complex-wavenumber and the complex-frequency/real-wavenumber pictures. These two viewpoints are shown to be self-consistent and both indicate that the PA condition occurs close to the ENZ frequency, at the crossover between bound and leaky behavior of one eigenmode of the system. At these special points, the field structure of the mode becomes indistinguishable from that of a totally absorbed incident plane wave [25]. The modal analysis also reveals the existence of a second PA condition for the ITO thin films at higher frequencies attributed to another crossover of the same eigenmode, which has been numerically verified using the three-medium Fresnel equation [26].

## 2. Experimental demonstration of perfect absorption in ultrathin, low permittivity films

Our experimental demonstration of perfect absorption in ultrathin, low permittivity films utilizes the three-medium structure shown in Fig. 1. It comprises a glass substrate (ambient dielectric, medium 1), an ultrathin ITO layer (medium 2), and an optically thick Ag capping layer (medium 3). For the reflectivity measurements, the optical waves are incident on the ITO film from the glass substrate side. The ITO films were deposited on boroaluminosilicate glass substrates using a 90 wt% $In_2O_3$ /10 wt% $SnO_2$ sputtering target at room temperature under a base vacuum pressure of $10^{-7}$ torr, followed by a 10 minute anneal in Ar gas at 700 ºC. Four samples were prepared with differing ITO thicknesses of $d$ = 24, 53, 88 and 137 nm. The permittivity functions of the four films in the spectral range of 0.3–2.5 µm were obtained through spectroscopic ellipsometry measurements, using a combination of Drude and Gaussian models. In addition, root-mean-square roughnesses of 0.2, 0.26, and 1.3 nm were measured using atomic force microscopy for the 24, 53 and 137 nm thick ITO films, respectively. Subsequently, a 200 nm layer of Ag was deposited on the ITO films using electron beam evaporation.

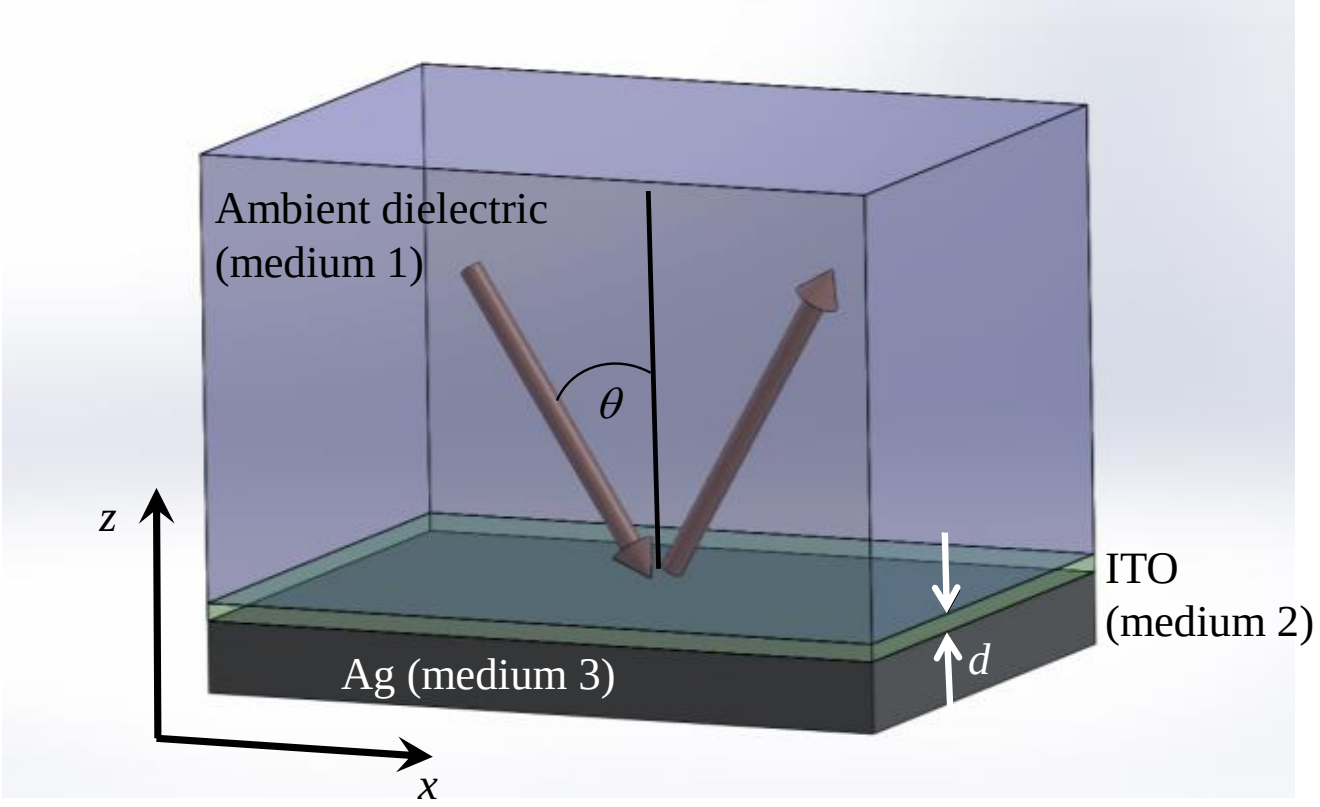

Fig. 1. Schematic depiction of a three-medium structure comprising an ambient dielectric (medium 1), a subwavelength ITO layer of thickness *d* (medium 2), and a metallic substrate (medium 3).

Since the Ag layer is opaque in the spectral range considered, the angular dependence of the reflection coefficient ( $\Gamma$ ) was measured using a spectroscopic ellipsometer (J.A. Woollam, Inc.), and the absorption ($A$) was computed as $A = 1 - |\Gamma|^2$. For these measurements, the incident beam was coupled to the sample through a glass prism that was index matched to the glass substrate of the sample, with the Ag surface facing outward in a fashion similar to previous measurements [27]. The incidence angle was varied in a stepwise manner, and a reflection spectrum spanning the range from 6000 cm$^{-1}$ to 12,000 cm$^{-1}$ (0.83 um to 1.67 um) was recorded for each angle.

The absorption spectra obtained at the incidence angles for which the maximum absorption was obtained are shown in Fig. 2, along with theoretical spectra computed using the three-medium Fresnel equation [26] (see Eq. (1) in Sec. 3) using the measured ITO permittivity and literature values for the permittivity of Ag [28]. A good agreement is observed between experimental and theoretical results. The maximum absorption values measured for the four samples are between 99.79-99.98%. We note that the instrumental limit of measuring zero reflectivity is 0.01%, and we estimate that other sources of error such as uncertainties in the determination of the permittivity functions, the layer thicknesses, and the wavelengths amount to an absorption uncertainty of 0.015%. A summary of the measured incidence angles, wavelengths, and values of maximum absorption for the four samples is presented in Table 1, and a plot of the maximum absorption angle versus reduced thickness ($d/\lambda_{PA}$) from both experiment and theory is reported in Fig. 3. Note that the wavelength at which perfect absorption was obtained increases with increasing sample thickness, while the perfect absorption angle decreases with increasing thickness.

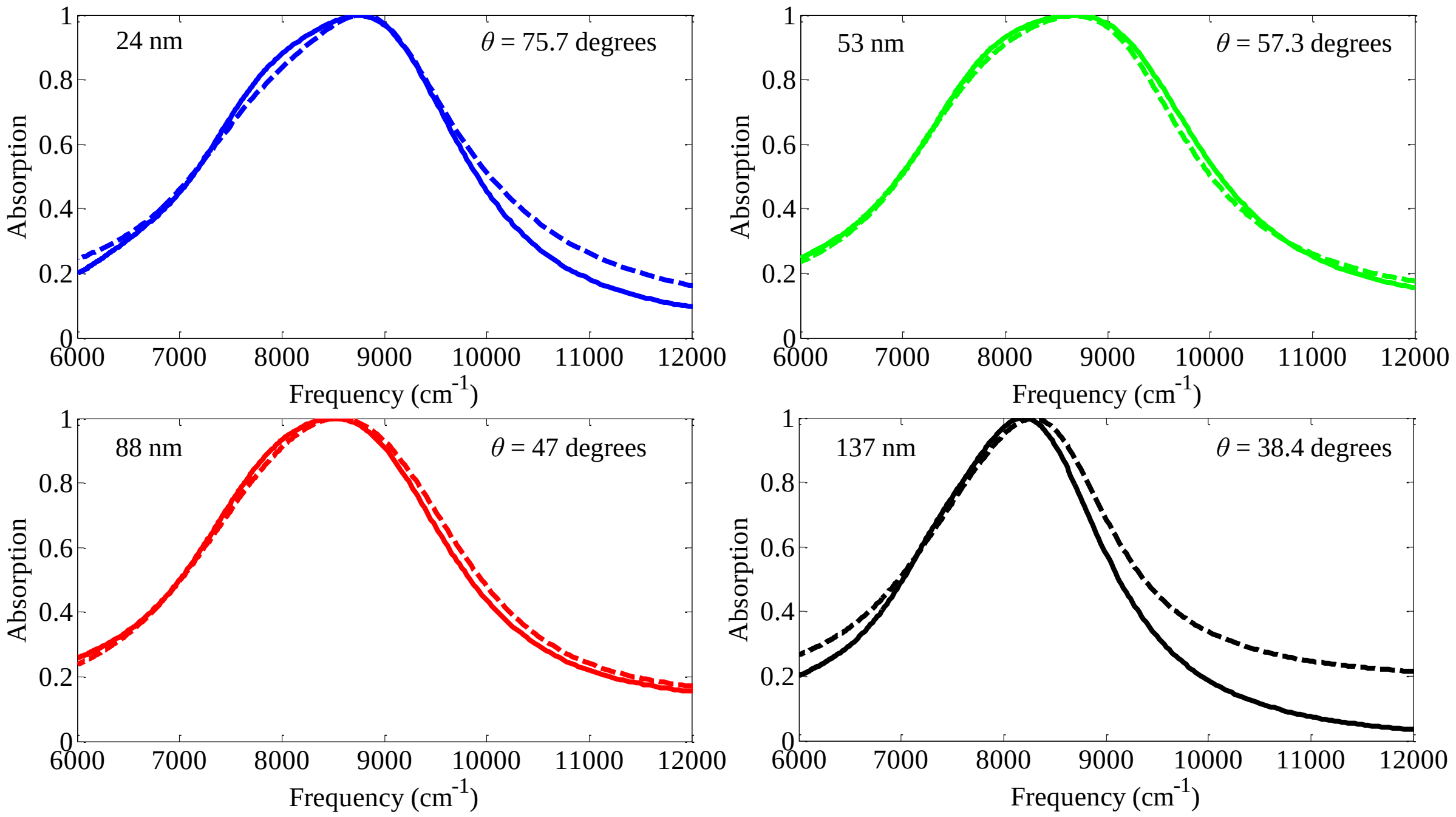

Fig. 2. Comparison of measured perfect absorption profiles (solid) and calculated profiles using three-medium Fresnel equation (dashed) for the four samples. The PA angle is explicitly indicated in each subplot.

Table 1. The experimental conditions for which perfect absorption was obtained for each of the four samples, showing the ITO thickness *d*, the ENZ wavelength $\lambda_{ENZ}$, the PA wavelength $\lambda_{PA}$, the PA incidence angle $\theta_{PA}$, the real ($\varepsilon_2'$) and imaginary ($\varepsilon_2''$) parts of the ITO permittivity at $\theta_{PA}$, and the experimental absorption value *A*. The values of $\lambda_{ENZ}$, and of $\varepsilon_2'$, and $\varepsilon_2''$ at $\lambda_{PA}$ for each sample were obtained using ellipsometry measurements for each ITO film.

| *d* (nm) | $\lambda_{ENZ}$ (μm) | $\lambda_{PA}$(μm) | $\theta_{PA}$ (degrees) | $\varepsilon_2'$ | $\varepsilon_2''$ | *A* (%) |
|---|---|---|---|---|---|---|
| 24 | 1.31 | 1.14 | 75.7 | 0.74 | 0.44 | 99.79 |
| 53 | 1.30 | 1.15 | 57.3 | 0.606 | 0.49 | 99.98 |
| 88 | 1.30 | 1.17 | 47 | 0.584 | 0.46 | 99.96 |
| 137 | 1.32 | 1.22 | 38.4 | 0.478 | 0.41 | 99.96 |

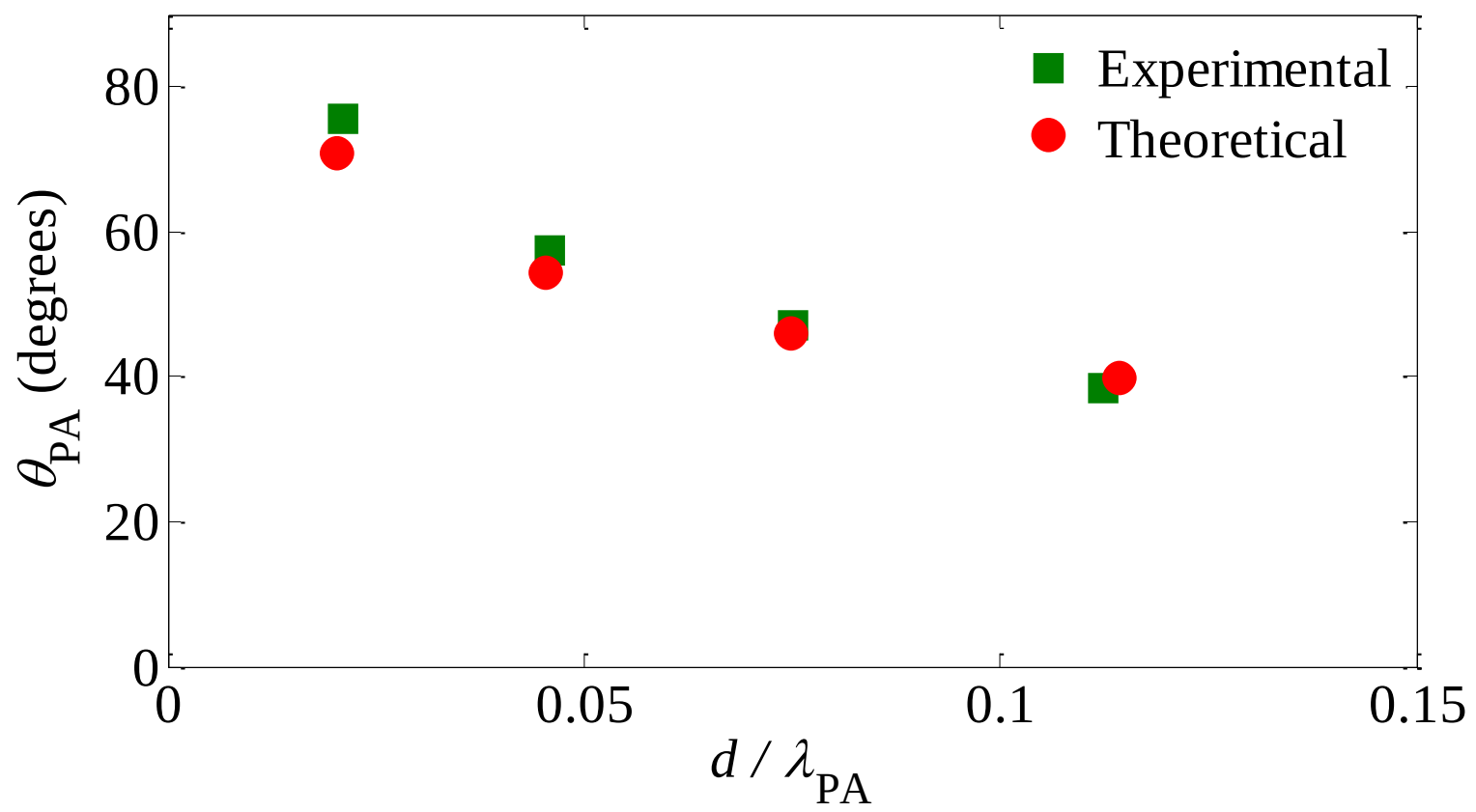

Fig. 3. Experimental (green squares) and theoretical (red circles) angles of the observed perfect absorption points versus the reduced ITO thickness.

A composite view of the experimental absorption spectra for the four samples is shown in the 1st column of Fig. 4. Each horizontal strip in this figure represents an absorption spectrum, measured at a particular incidence angle. The 2nd column of Fig. 4 displays the theoretical results computed using the three-medium Fresnel equation [26]. These calculations are in good agreement with experimental results. On this figure, the perfect absorption point closest to the ENZ frequency is marked by a magenta cross (low-frequency PA). Careful inspection of the theoretical absorption maps shows the presence of a second PA condition that is beyond our experimental measurement range and is marked by a cyan square (high-frequency PA). We also show on the theoretical absorption maps a solid black curve which is the locus of points depicting the behavior of the propagation constant (i.e. the real part of the complex modal wavenumber) of one of the electromagnetic eigenmodes of the system shown in Sec. 4. Note that the trajectory of this particular eigenmode precisely crosses both perfect absorption points (a deeper discussion of the modal dispersion will be provided in the next sections).

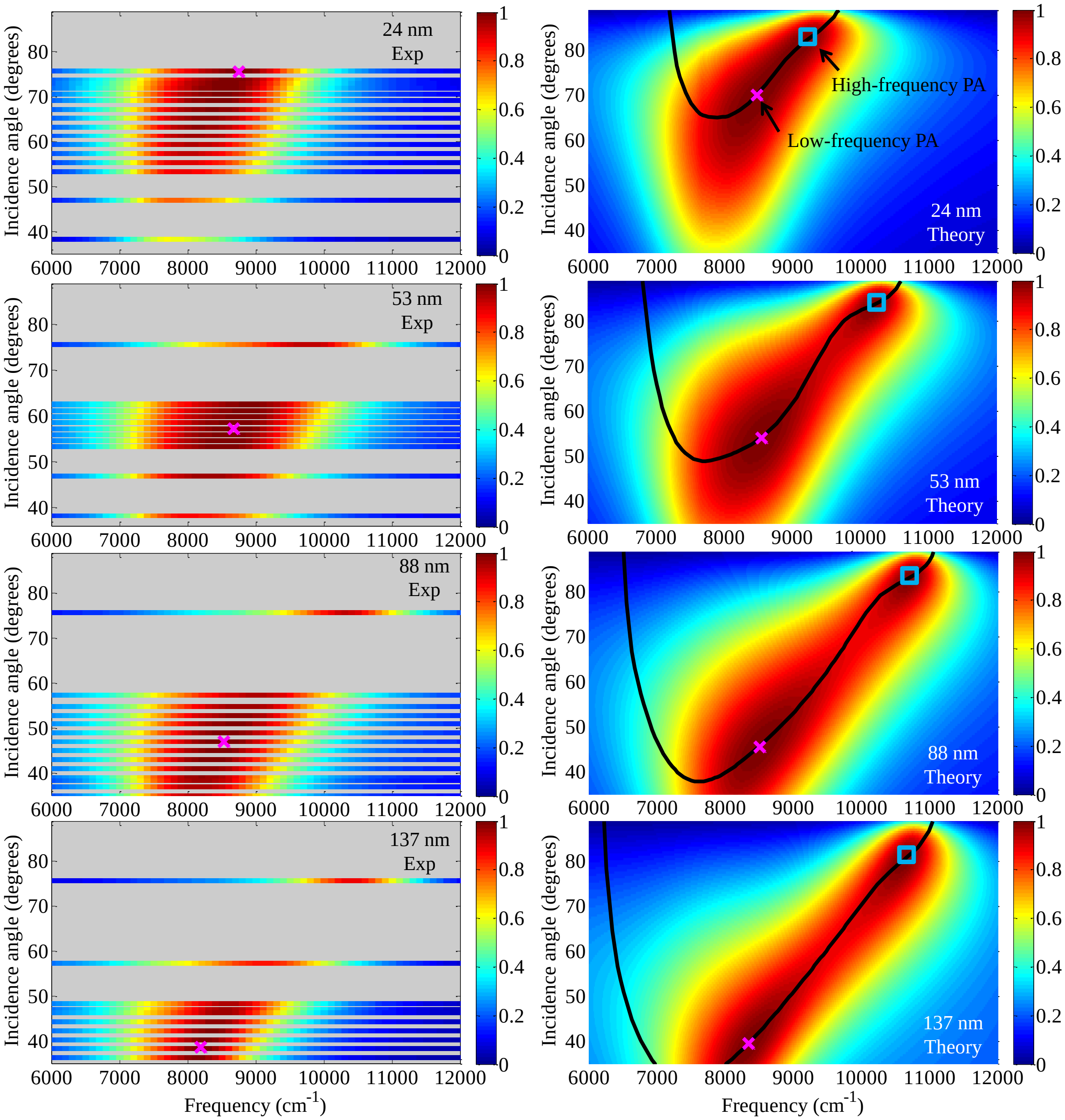

Fig. 4. Comparison of measured angular and frequency dependence absorption profile of the four samples comprising ITO with $d$ = 24, 53, 88 and 137 nm on Ag. The first column displays the raw experimental spectra at several discrete angles and the second column shows results calculated from Eq. (1) using the experimental permittivity function for ITO and literature values for the permittivity of Ag [28]. The calculated absorption maps show two perfect absorption conditions that are indicated with magenta crosses and cyan squares. On these maps we superimpose a solid black curve which is the locus of points depicting the behavior of the propagation constant of one of the electromagnetic eigenmodes of the system shown in Sec. 4.

To understand the spatial dependence of the absorption, we performed full-wave finite-difference time-domain simulations (FDTD Lumerical Solutions) using the perfect absorption conditions for the 53 nm sample as a representative case. In Fig. 5(a), we show the spatial profile of the absorption per unit length in units of m$^{-1}$ (proportional to $\varepsilon_0\varepsilon_2''|\mathbf{E}|^2/P_{\text{inc}}$ times the unit cell area used in simulation for convenience, with $P_{\text{inc}}$ the incident power) computed using the simulated total electric field $\mathbf{E}$. Inspection of Fig. 5(a) reveals that most of the energy is absorbed within the ITO layer. More quantitatively, numerical integration of the absorption per unit length in Fig. 5(a) along the thickness of the ITO film and the Ag region reveals that ~99% of the incident energy is absorbed within the deeply subwavelength ITO film and the remaining power dissipation (~1%) occurs in the Ag. Furthermore, the absorption per unit length is maximum at the glass/ITO interface and drops monotonically throughout the thickness of the film as shown in [24]. Compared to the ITO single pass absorption of about 5%, the absorption enhancement factor is about 20. Figure 5(b) shows the enhancements of the *z*-component of the electric field that are observed within the ITO layer in the vicinity of the ENZ frequency. Note that, as also discussed in [29] for a subwavelength ENZ layer sandwiched between two silica domains, the maximum field enhancement does not occur at $\mathrm{Re}(\varepsilon)=0$ (indicated by a solid white line) where an enhancement of $|E_z|^2 \approx 4$ is observed (in the simulations, the total incident field strength is 1), but rather at a larger frequency. At the perfect absorption frequency (indicated by a dashed white line), which is slightly larger than the $\mathrm{Re}(\varepsilon)=0$ frequency, the enhancement is approximately $|E_z|^2 \approx 5$. While a maximum enhancement of $|E_z|^2 \approx 7$ is observed at a somewhat higher frequency.

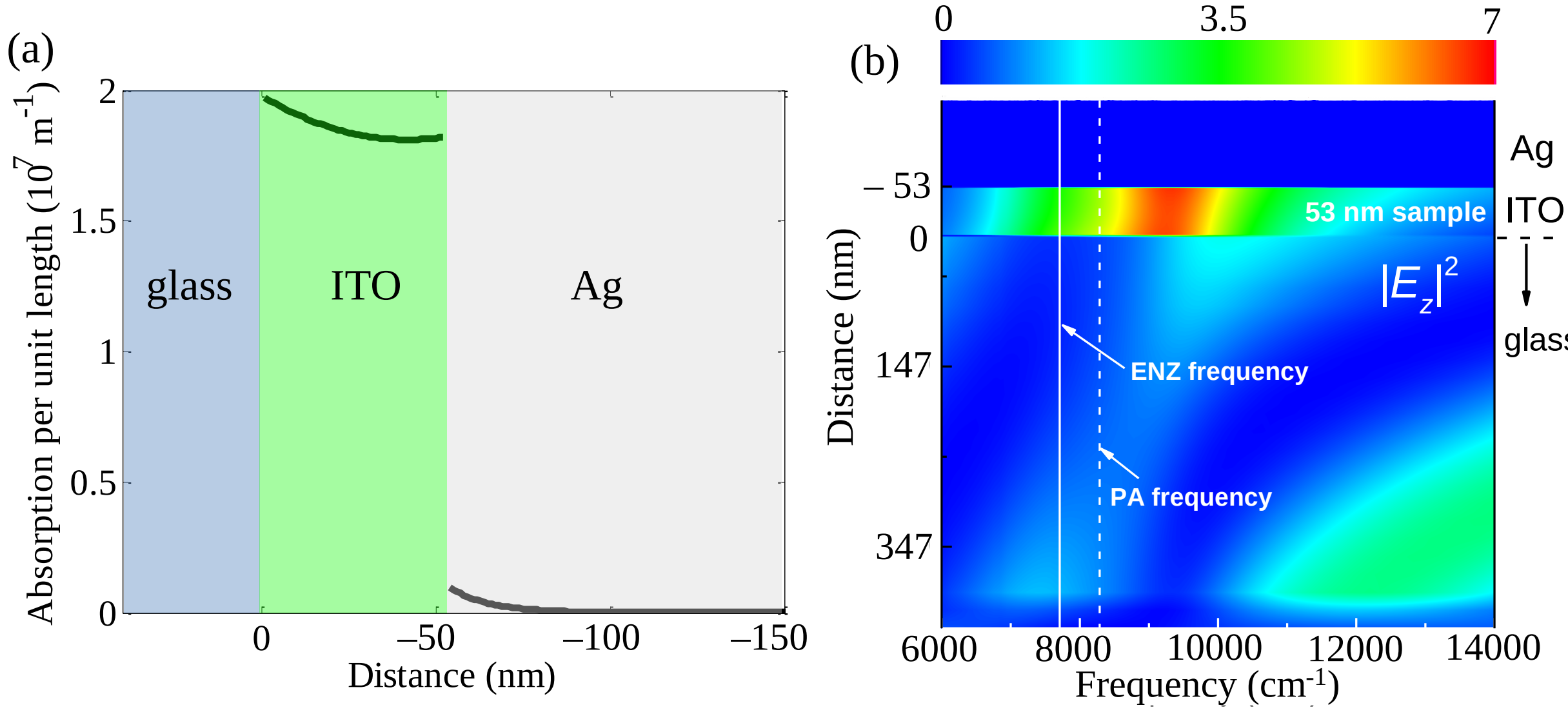

Fig. 5. (a) The spatial profile of the absorption per unit length in units of $\mathrm{m}^{-1}$ for the 53 nm sample at the PA frequency. Most of the energy is absorbed within the ITO layer. (b) The spatial distribution of the *z*-component of the electric field intensity versus frequency, showing ENZ related field intensity enhancements. The maximum field enhancement is at a larger frequency than the one where $\mathrm{Re}(\varepsilon)=0$, in agreement with previous investigations [29].

## 3. Perfect absorption in subwavelength films – simple theoretical framework

In this section we present a simple theoretical analysis of the conditions required to achieve perfect absorption in subwavelength films. This will be followed in subsequent sections by more rigorous numerical analyses of the perfect absorption phenomenon. Consider the three-medium structure shown in Fig. 1. Media 1, 2, and 3 are the ambient dielectric, the low permittivity material, and metallic substrate with relative permittivities of $\varepsilon_1$, $\varepsilon_2$, and $\varepsilon_3$, respectively. The reflection coefficient $\Gamma$ of such a structure is given by the standard three-medium layer Fresnel equation [26]

$$\Gamma = \frac{r_{12} + r_{23} e^{2i\phi}}{1 + r_{12} r_{23} e^{2i\phi}} \tag{1}$$

where $r_{ij}$ is the reflection coefficient of the *ij* interface, with $i, j = 1,2,3$, $d$ is the layer thickness, and $\phi = k_{z2} d$ is the single pass phase shift across the low permittivity layer. The longitudinal wavenumber, $k_{z2}$, is related to the component of the wavenumber parallel to the interface by the relation $k_{z2}^2 = \varepsilon_2 k_0^2 - k_x^2$, where $k_x = k_0 \sqrt{\varepsilon_1} \sin\theta$, and $k_0$ and $\theta$ are the free-space wavenumber and the (real) incident angle, respectively.

For simplicity, we will begin by assuming medium 3 to be a perfect electric conductor (PEC), which implies that $r_{23} = -1$. In this case, we see from Eq. (1) that the reflection coefficient will vanish when $r_{12} = e^{2i\phi}$ (leading to the PA condition). If medium 2 exhibits loss, then physical solutions corresponding to $\Gamma = 0$ can be found. Assuming *p*-polarized incidence and that the thickness of medium 2 is deeply subwavelength (i.e., $\phi$ is small), the condition $r_{12} = e^{2i\phi}$ can be rewritten as [24]:

$$\frac{k_{z2}^2 d}{\varepsilon_2} = i\frac{k_{z1}}{\varepsilon_1}. \tag{2}$$

Using $k_{z2}^2 = \varepsilon_2 k_0^2 - k_x^2$ and noting that $\varepsilon_2$ is a complex quantity, we can equate the real and imaginary parts of Eq. (2), to obtain after some manipulation the PA conditions:

$$\sqrt{\varepsilon_1}\sin\theta_{PA} = \sqrt{\frac{\varepsilon_2^{'2} + \varepsilon_2^{''2}}{\varepsilon_2^{'}}} \tag{3}$$

$$\frac{2\pi d_{PA}}{\lambda_{PA}} = \frac{\varepsilon_2^{'}\cos\theta_{PA}}{\sqrt{\varepsilon_1}\varepsilon_2^{''}} \tag{4}$$

We can further combine Eqs. (3) and (4) to obtain

$$\frac{2\pi d_{PA}}{\lambda_{PA}} = \left[\frac{\varepsilon_2^{'\,2} + \varepsilon_2^{''2}}{\varepsilon_1^{3/2}\varepsilon_2^{''}}\right]\frac{1}{\tan\theta_{PA}\sin\theta_{PA}}, \tag{5}$$

which is identical to the expression obtained using different considerations in Ref. [23]. We note that, to be consistent with the choice of small $\phi$ that led to Eq. (2), the acceptable choices for the parameters in Eqs. (4) and (5) must lead to a deep subwavelength film thickness. Thus, if the permittivity functions are such that the prefactor of Eq. 5 is not small, then extremely oblique angles must be used to maintain the thin film condition. Furthermore, we note that the thickness $d_{PA}$ will approach zero if simultaneously $\varepsilon_2^{'}$ and $\varepsilon_2^{''}$ approach zero in such a manner that the prefactor of Eq. 5 approaches zero (i.e. if $\varepsilon_2^{'}$ approaches zero faster than $\varepsilon_2^{''}$).

Inspection of Eqs. (3) and (4) reveals the conditions that the complex permittivity $\varepsilon_2$ must satisfy if perfect absorption is to be achieved in the deep subwavelength limit. From Eq. (4) we see that perfect absorption is possible only when $\varepsilon_2^{'} > 0$. Therefore, the frequency at which perfect absorption occurs is larger than the ENZ frequency $f_{ENZ}$, defined as $\varepsilon_2^{'}(f_{ENZ}) = 0$, in agreement with what indicated in Fig. 5(b). We obtain two further conditions the permittivity must satisfy by requiring that the incidence angle in Eq. (3) remains real:

$$\varepsilon_2^{''} < \frac{\varepsilon_1}{2} \tag{6}$$

$$\frac{\varepsilon_1}{2} - \sqrt{\frac{\varepsilon_1^2}{4} - \varepsilon_2^{''2}} \le \varepsilon_2^{'} \le \frac{\varepsilon_1}{2} + \sqrt{\frac{\varepsilon_1^2}{4} - \varepsilon_2^{''2}} \tag{7}$$

Figure 6 shows the two bounds given by Eqs. (6) and (7). The shaded region between the red and blue lines indicates an area where perfect absorption solutions can be found. Indeed, all the four samples described in Sec. 2 are within the shaded region, as indicated by the green squares. We emphasize that the solution shown in Fig. 6 was obtained in the PEC approximation and therefore only serves as a starting point for assessing the possibility of perfect absorption.

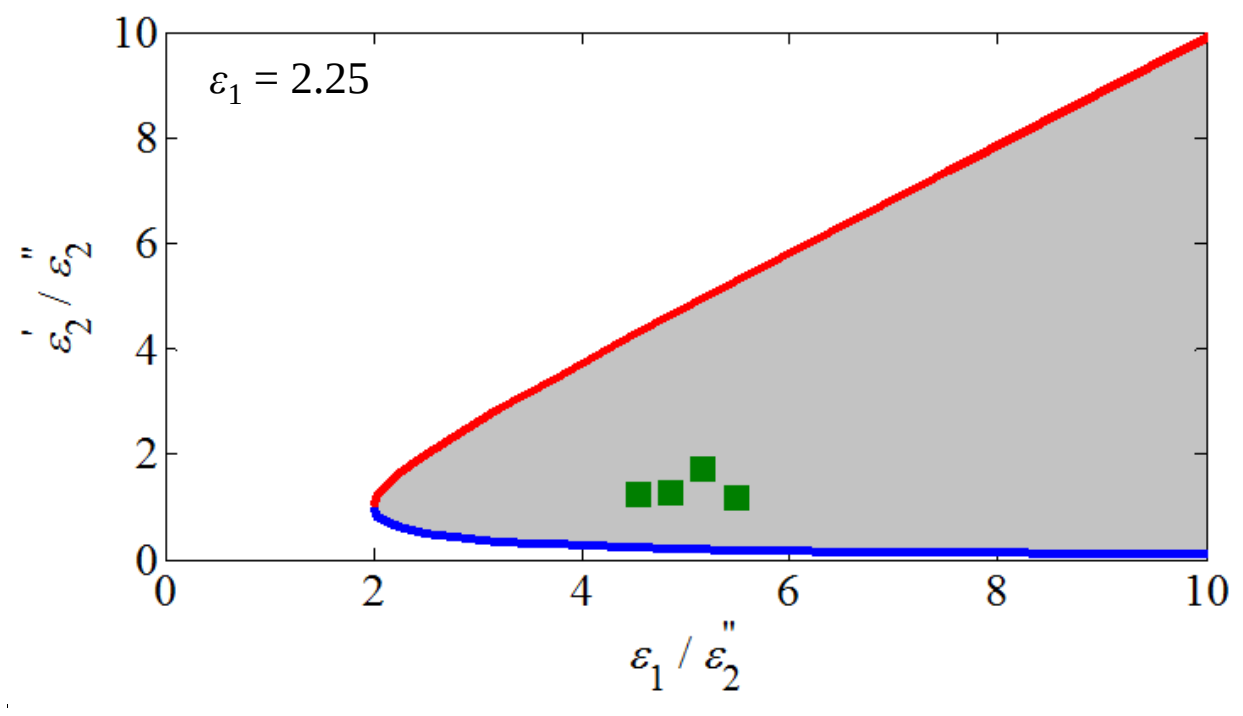

Fig. 6. Upper (red) and lower (blue) bound of $\varepsilon_2' / \varepsilon_2''$ given by Eqs. (6) and (7). The grey shaded region indicates the area where perfect absorption solutions can be found. The parameters of the four ITO samples studied here are shown in green squares. All the analyzed designs fall within the PA region.

## 4. Origin of perfect absorption 1 – Real-frequency/Complex-wavenumber modal dispersion diagrams

To unravel the physical origin of the perfect absorption conditions observed in Sec. 2, we investigate the eigenmodes of the three-medium structure. While other theoretical methods such as critical coupling theory [30-32] or partial wave sums [19] could be used to analyze this system, we find the ability to locate the eigenmodes in the complex-wavenumber or complex-frequency plane extremely informative. Of primary interest are modes that lie close to the real frequency or real wavenumber axes, since these will couple most strongly to the incoming plane waves [33, 34]. Furthermore, by monitoring the modal dispersion diagrams variation as a function of layer thickness or constituent material permittivities we can systematically modify the structure to move the modes to desired locations [15, 17]. In this section, we compute the real-frequency/complex-wavenumber modal dispersion diagrams of modes supported by the three-medium structure, while in the next section we consider the complementary calculation of the real-wavenumber/complex-frequency dispersion diagrams.

Assuming a reference plane at the center of the ITO layer, the real-frequency/complex-wavenumber dispersion diagrams can be obtained by finding the complex roots of the equation $Z_t = Z_u + Z_d = 0$ for *p*-polarized waves, where $Z_u = \frac{A_2 Z_1 + B_2}{C_2 Z_1 + D_2}$ is the wave impedance of medium 1 transferred to the ITO center, $Z_d = \frac{A_2 Z_3 + B_2}{C_2 Z_3 + D_2}$ is the wave impedance of medium 3 transferred to the ITO center, and $A_2$, $B_2$, $C_2$, and $D_2$ are the ABCD parameters of the ITO layer [35]. The longitudinal wavenumber $k_{zn}$ in the *n*th media can then be computed using $k_{zn}^2 = k_n^2 - k_x^2$. Then, to find the eigenmodes, we fix the (real) frequency and find the transverse wavenumber ($k_x = \beta_x + i\alpha_x$) by calculating the roots of $Z_t = Z_u + Z_d = 0$. Note that for *p*-polarized waves, the characteristic impedance of each layer becomes $Z_n = \frac{k_{zn}}{\omega \varepsilon_0 \varepsilon_n}$ and $k_n^2 = k_0^2 \varepsilon_n$, where $\varepsilon_0$ is the absolute permittivity of free space and $\omega$ is the angular frequency. [We note that modal dispersion diagrams can equivalently be determined by finding the complex roots of the denominator of Eq. (1).] Application of this procedure leads to four eigenmodes for each frequency, with each eigenmode being associated with a particular choice of signs for $\pm k_{z1}$ and $\pm k_{z3}$ [25]. In general, these four modes correspond to bound or leaky waves of the structure, with incoming or outgoing wave propagation in the bounding half spaces (i.e., media 1 and 3). In the following we shall retain only the particular eigenmode that we have found to be associated with the perfect absorption process.

Application of this numerical procedure leads to the dispersion diagrams depicted in Fig. 7 for the four samples investigated in this paper. As mentioned above, only one mode is retained in Fig. 7 since it plays a key role in explaining both perfect absorption conditions observed in Fig. 4. The first column of Fig. 7 shows the dependence of the propagation constant $\beta_x$ (i.e., the real part of the complex modal wavenumber) on frequency, while the second column shows the frequency dependence of the attenuation constant $\alpha_x$ (i.e., the imaginary part of the complex modal wavenumber). The third column maps the dispersion diagram in the complex $\beta_x - \alpha_x$ plane. Inspection of the dispersion curves reveals that the amplitude of the switch-back region and the attenuation constant both increase with increasing thickness. Furthermore, we observe that the attenuation constant crosses zero twice for each film thickness, a very important condition to explain perfect absorption that will be discussed next. This is equivalent to the mode trajectory crossing the real axis of the complex plane twice, as is seen in the third column of Fig. 7.

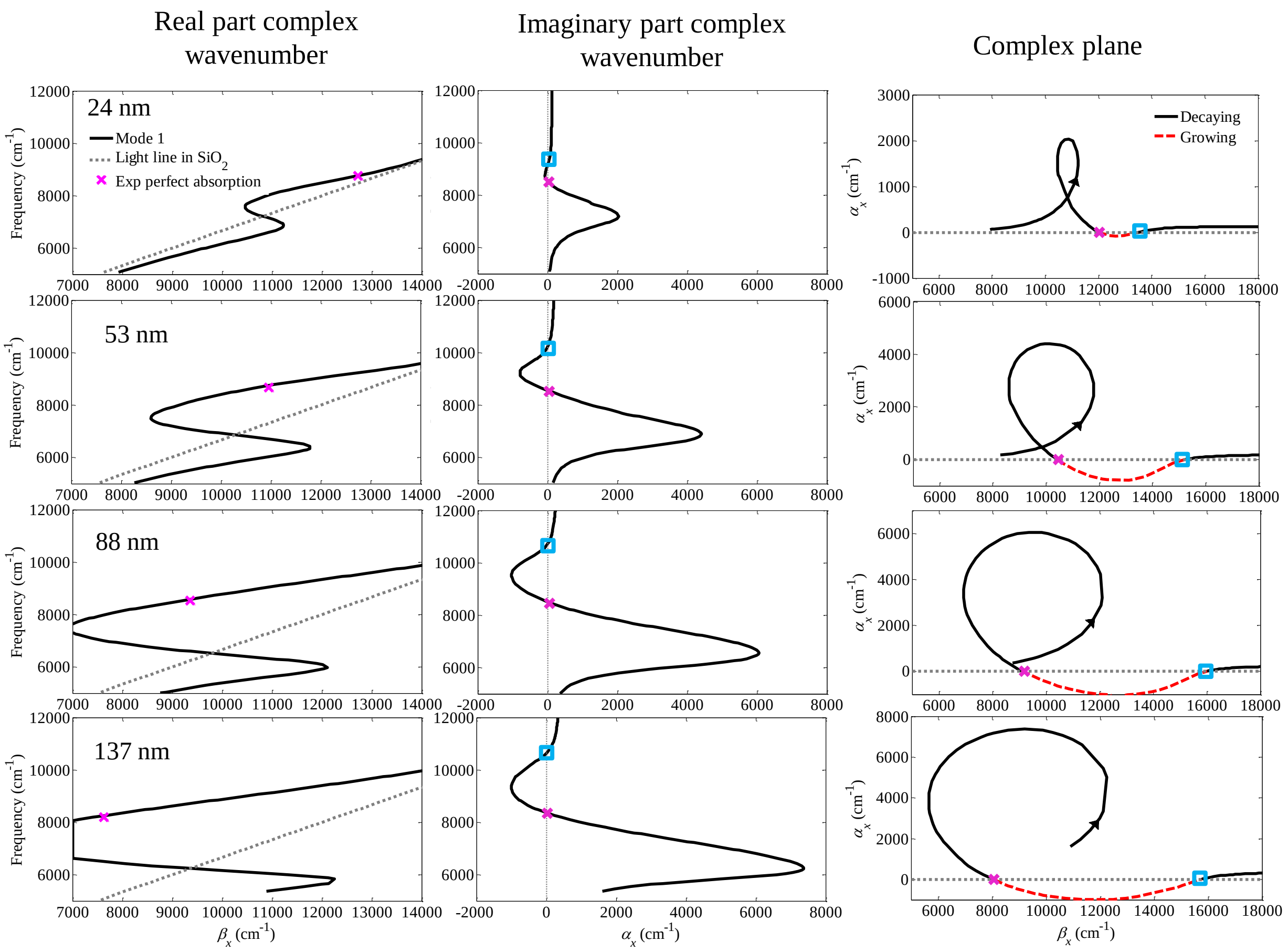

Fig. 7. Real-frequency/complex-wavenumber modal dispersion diagrams pertaining to the four samples comprising ITO with $d$ = 24, 53, 88 and 137 nm. The first column shows the frequency/propagation constant dispersion, the second column shows the frequency/attenuation constant dispersion, and the third column shows the dispersion diagram in the complex wavenumber plane. Portions of these dispersions shown in column 1 were previously shown as black lines in column 2 of Fig. 4. In the first column, the experimental data points representing perfect absorption of the four samples are shown as magenta crosses. The dotted gray line represents the light line in medium 1. In the second and third columns, the two perfect absorption conditions are marked with magenta crosses and cyan squares, as done in Fig. 4. Moreover, in the third column, the arrows indicate the direction of increasing frequency.

The mode shown in Fig. 7 is obtained with a sign choice for $k_{z1}$ that leads to incoming wave character in medium 1 for all frequencies. At low frequencies this mode is bound (i.e. exponentially decaying in medium 1, solid black in third column of Fig. 7), but with increasing frequency the trajectory of the mode in the complex $\beta_x - \alpha_x$ plane crosses the real axis and the mode becomes unbound (i.e. leaky, dashed red in third column of Fig. 7). As the frequency continues to increase, the mode re-crosses the real axis and once again becomes bound (solid black in third column of Fig. 7). *The points where this mode crosses the real axis are the perfect absorption points*, marked with magenta crosses and cyan squares for convenience in both second and third columns of Fig. 7. At these points, the field in medium 1 is neither growing nor decaying, but rather it is the field of an incoming plane wave. Thus, at these points the field structure of this mode, which is an exact solution to Maxwell's equations, is indistinguishable from the field structure of an incoming and totally absorbed plane wave — *perfect absorption*. Thus, at these points (and in their close vicinity) a plane wave can phase match to the mode in the manner described in [33, 34]. Note that the sign choices required for the physically measured reflectivity ( $+k_{z1}$ and $+k_{z3}$) differ from the sign choices for the mode whose dispersion is shown in Fig. 7 ( $-k_{z1}$ and $+k_{z3}$) by the choice of sign for $k_{z1}$. As was shown in Ref. [14], $\Gamma(k_{z1}) = \Gamma^{-1}(-k_{z1})$, so that the pole of the reflection coefficient for the mode of interest becomes a zero of the experimental reflectivity, further justifying the overlap of the modal dispersion with the PA points in Fig. 4. The existence of *two* perfect absorption points is fully consistent with the reflectivity maps shown in Fig. 4 which were calculated using the same parameters (the zero crossing points are again marked with magenta crosses and cyan squares). Due to experimental limitations of the incidence angle, we are only able to obtain experimental verification of the perfect absorption point near the frequency of 8500 cm$^{-1}$.

## 5. Origin of perfect absorption 2 – Complex-frequency/Real-wavenumber modal dispersion diagrams

In this section we present the complementary analysis of the perfect absorption process using complex-frequency/real-wavenumber modal dispersion diagrams. These dispersion diagrams can be obtained by using a method similar to that used in Sec. 4 by finding the complex frequency roots of the denominator of Eq. (1) for real $k_x$. As done in Sec. 4, we will retain only the relevant mode, computed using the sign choices $-k_{z1}$ and $+k_{z3}$, of the four possible modes that can be computed (the dispersion diagrams of all the possible sign choices are reported in the Appendix for one sample). In these calculations the dielectric functions of the ITO samples were approximated with complex Drude models since frequency-dependent permittivity values are required for complex frequencies. In Fig. 8 we show a complex frequency map of the natural log of the reflectivity calculated from Eq. (1) at the low-frequency perfect absorption angle of each sample. For arbitrary angles, the zero reflectivity point will not lie on the real axis. However, for angles which correspond to the experimentally determined perfect absorption angles, the perfect absorption point can be seen as a small blue region ("blue dot") on the real axis (see Fig. 8). Superimposed on the reflectivity map is the trajectory of the mode that is the complex frequency analog of the mode described in Sec. 4. Once again, we observe that the mode trajectory crosses the real axis in two locations, and for the maps shown here, the low frequency crossing directly coincides with the zero of the reflectivity. [If the reflectivity maps were calculated for the conditions corresponding to the high-frequency perfect absorption, then the blue dots would be located at the points where the trajectories cross the real axis at higher frequencies (see Appendix).] In agreement with the results of Sec. 4, two perfect absorption points are observed, and the wavenumber and frequency values of the perfect absorption points (which are real at these locations) agree with the corresponding values obtained from the complex wavenumber analysis. Furthermore, in agreement with the results of Sec. 4, the locations where the mode trajectory crosses the real axis correspond to points at which the field of the mode corresponds to an incoming plane wave in medium 1, and the overall field profile is indistinguishable from the fields occurring for perfect absorption.

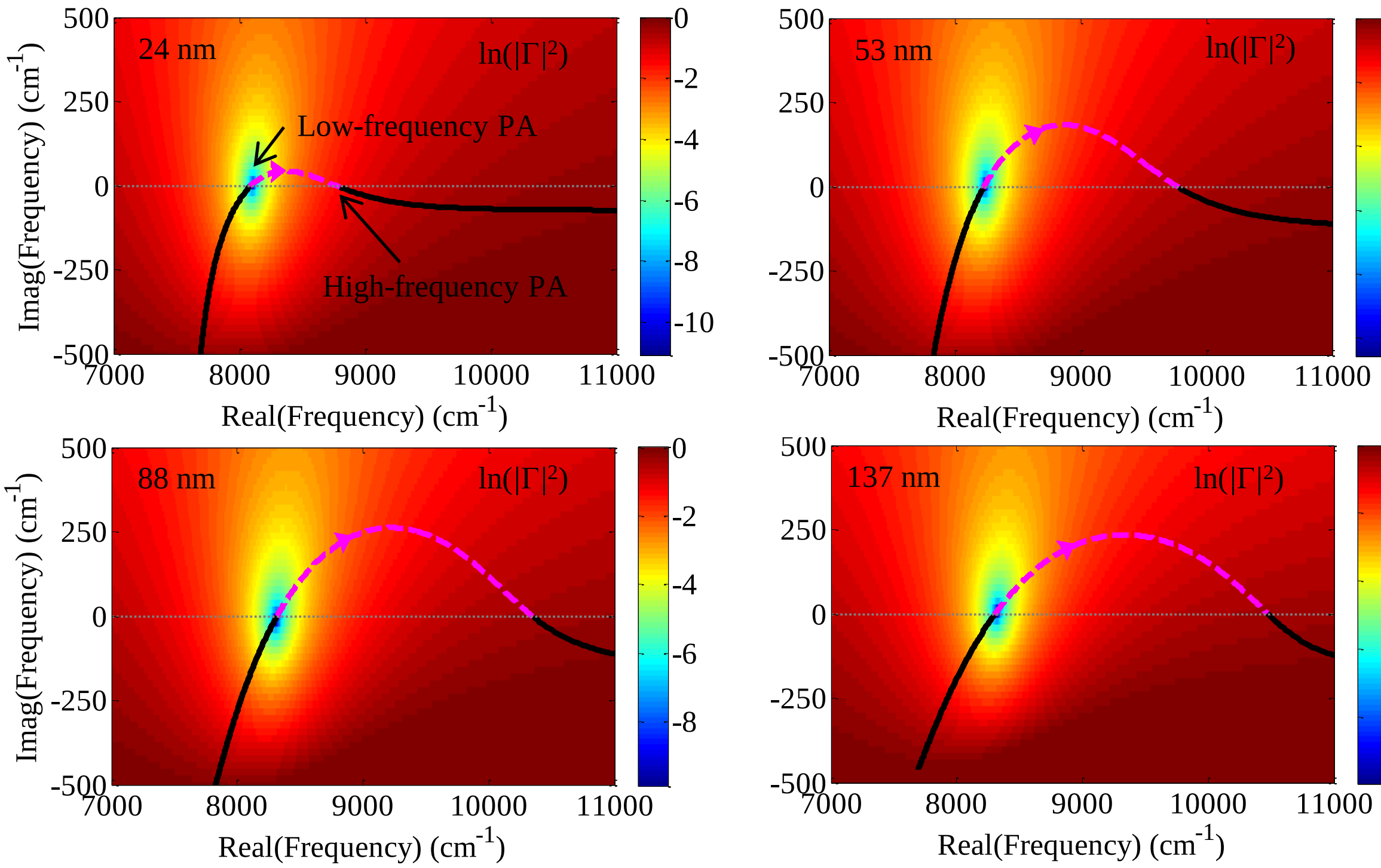

Fig. 8. Complex frequency maps of the natural logarithm of the reflectivity for the four samples computed from Eq. (1) at the PA incident angle. The zero reflectivity point (blue dot) represents the perfect absorption point. The solid and dashed lines are the dispersion diagrams of eigenmodes of the three-medium structure for increasing incidence angle indicated by the arrows. A solid line indicates a mode decaying in both medium 1 and 3, a dashed line indicates a mode decaying in medium 3 and growing in medium 1. Note that as observed in Fig. 4, we again achieve two perfect absorption points around 8500 and 10000 cm$^{-1}$ in correspondence of crossing of the Im(ω)=0 axis.

## 6. Conclusion

We have provided an experimental demonstration that single beam directional perfect absorption can be achieved using deep subwavelength films of low-permittivity materials on metal substrates. Furthermore, we have shown that in the perfect absorption process, essentially all of the incoming energy is absorbed in the thin film material and not in the metallic substrate. While the results obtained in this paper specifically related to ITO thin films, it is clear that similar behavior should be expected for any material with similar dielectric properties, in any frequency range. For example, the perfect absorption behavior described in this paper is fully consistent with the results of Refs. [22, 23], and explains why the absorption observed in that work became stronger as the frequency was increased above the ENZ frequency. Thus, our results further expand the list of possible material candidates for use in perfect absorption applications, to include: other diluted plasmas such as other conducting oxides, heavily doped semiconductors; restrahlen materials just above the longitudinal optical phonon frequencies; and correlated electron materials such as vanadium dioxide.

We have provided simple analytical formulas that relate material parameters to regions where perfect absorption solutions can be found, and therefore can be used as a starting point for assessing the possibility of perfect absorption. We have further analyzed the electromagnetic eigenmodes of the layered structure from two complementary viewpoints and have showed that the perfect absorption arises due to the behavior of a particular eigenmode of the system. Although in general this mode exhibits an unphysical field profile, it is nevertheless an exact solution of Maxwell's equations. Importantly, the mode field profile becomes identical to that of a totally absorbed incoming plane wave at the crossover between bound and leaky behavior, a behavior that was recognized in [25]. Hence, the perfect absorption process is also an exact solution to Maxwell's equations. Identification of the eigenmode responsible for perfect absorption will allow new designs to be obtained by adjusting material and structural parameters to force the trajectory of this mode to cross the real axis at desired locations. Furthermore, this procedure is general and can also be employed in more complex multilayer systems. Interestingly, our analysis has revealed the existence of a second set of perfect absorption points which are outside the accessible measurement range of our instrumentation. Thus, these results provide a new path for the design of custom absorber materials that do not rely on surface patterning or texturing and, hence, will be more appealing for device applications.

## Acknowledgements

The authors acknowledge stimulating discussions with Tom Tiwald, Mathias Schubert, Xiaodong Yang, and Zongfu Yu. This work was supported by the U.S. Department of Energy, Office of Basic Energy Sciences, Division of Materials Sciences and Engineering and performed, in part, at the Center for Integrated Nanotechnologies, an Office of Science User Facility operated for the U.S. Department of Energy (DOE) Office of Science. Sandia National Laboratories is a multi-program laboratory managed and operated by Sandia Corporation, a wholly owned subsidiary of Lockheed Martin Corporation, for the U.S. Department of Energy's National Nuclear Security Administration under contract DE-AC04-94AL85000.

## Appendix: Detailed complex-frequency/real-wavenumber modal dispersion diagrams

In this Appendix we use the method described in Secs. 4 and 5 to compute the complex-frequency/real-wavenumber modal dispersion diagrams of all the modes supported by the three-medium structure for the 53 nm sample. Figure 9 shows four complex frequency maps of the natural log of the reflectivity corresponding to four different incidence angles. As mentioned in Sec. 5, for arbitrary angles the zero reflectivity point ("blue dot") will not lie on the real axis, but does lie on the real axis at the PA angle. This can be clearly observed in Fig. 9 where the position of the "blue dot" depicting the reflectivity minimum moves in the complex frequency plane as the incidence angle is varied. Superimposed on the reflectivity maps are the trajectories of the modes (without distinguishing between bound or leaky for simplicity and clarity of presentation) computed with the following sign choices: 1) $+k_{z1}$ and $+k_{z3}$ (dashed white); 2) $-k_{z1}$ and $+k_{z3}$ (dotted green); 3) $+k_{z1}$ and $-k_{z3}$ (solid blue); and 4) $-k_{z1}$ and $-k_{z3}$ (dash-dotted black). We observe that only the trajectory of the dotted green mode, computed with the sign choice $-k_{z1}$ and $+k_{z3}$, follows the locus of the reflectivity minima for various incidence angles, which clearly demonstrates that this particular eigenmode is responsible for the perfect absorption process.

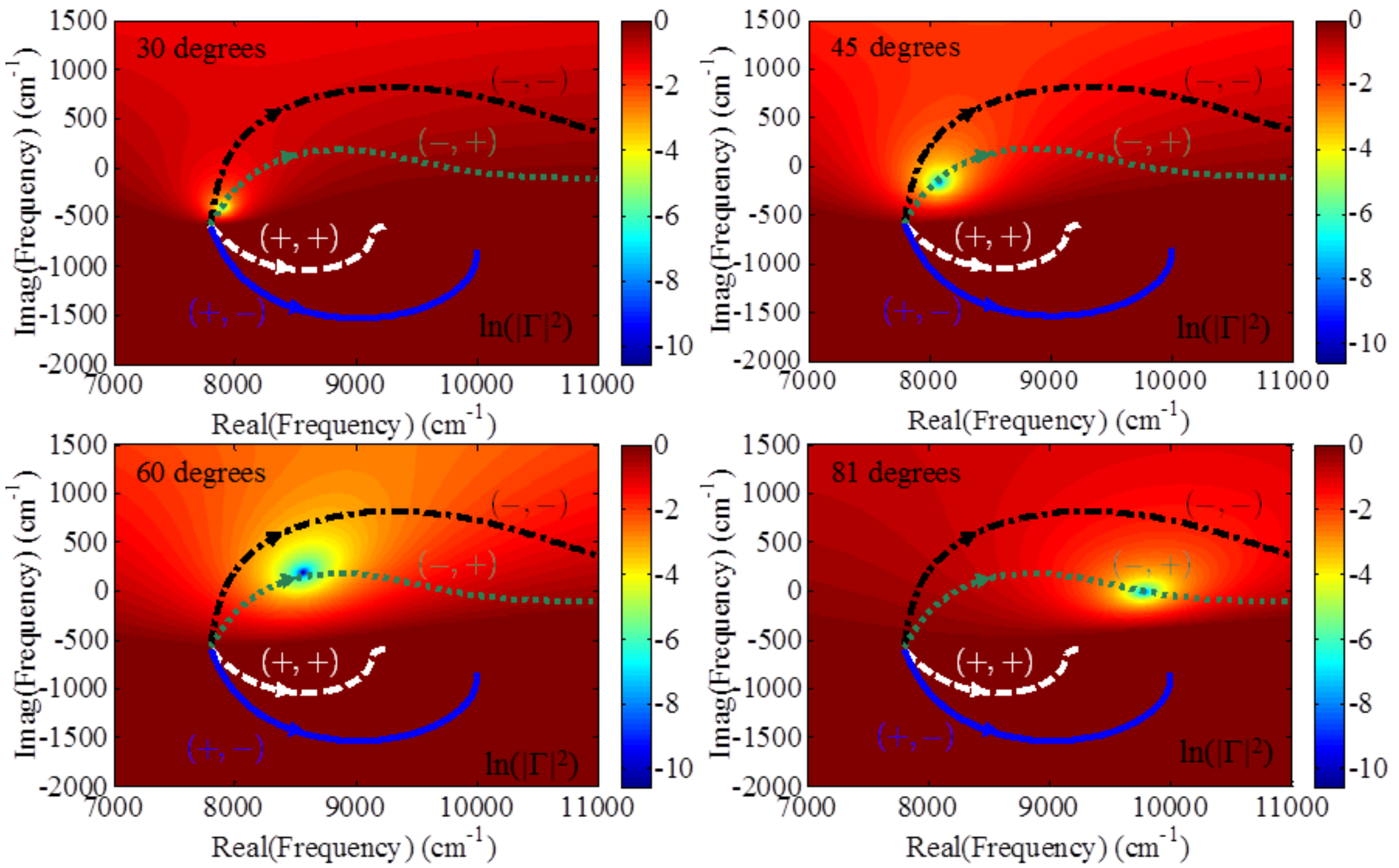

Fig. 9. Complex frequency maps of the natural logarithm of the reflectivity for the 53-nm-thick sample computed from Eq. (1), for four different incidence angles. Note that only the trajectory of the eigenmode computed using the sign choice $-k_{z1}$ and $+k_{z3}$ follows the reflectivity minima for different incidence angles, further confirming that this eigenmode is associated with the perfect absorption process.